\documentclass{ws-ijqi}

\begin{document}

\markboth{Tien D. Kieu}
{Transcending Turing computability}

\catchline{}{}{}{}{}

\title{An anatomy of a quantum adiabatic algorithm that\\transcends 
the Turing computability}

\author{Tien D. Kieu}
\address{Centre for Atom Optics and Ultrafast Spectroscopy, Swinburne 
University of Technology, Hawthorn, VIC 3122, Australia\\kieu@swin.edu.au}


\maketitle

\begin{history}
\received{(DAY MONTH YEAR)}
\revised{(DAY MONTH YEAR)}
\end{history}

\begin{abstract}
We give an update on a quantum adiabatic algorithm for the Turing noncomputable 
Hilbert's tenth problem, and briefly go over some relevant issues 
and misleading objections to the algorithm.
\end{abstract}

\keywords{Quantum adiabatic computation; Turing computability; Hilbert's 
tenth problem}

\section{Introduction}
Hilbert's tenth problem~\cite{Hilbert's} asks for a {\it single} procedure/algorithm 
to systematically determine if any given Diophantine equation has some positive integer 
solution or not.  This problem has now been proved to be Turing noncomputable, 
indirectly through
an equivalence mapping to the noncomputable Turing halting problem~\cite{recursive}.

Nevertheless, we have proposed and argued (with more and more details provided) in a series of 
papers~\cite{kieuIJTP,kieuCP,kieuPRS,kieuOrlando,kieuFull}
for an algorithm for  Hilbert's tenth problem based on quantum adiabatic 
computation~\cite{QAC}.  
Among the many valid concerns about our algorithm there are also some misleading 
objections which are still being spread despite their falsehood.
In this short note we will give an updated overall picture of the algorithm and
go through, with pointers provided for further discussions 
elsewhere, some of the valid concerns as well as the misleading objections to dispel
the entrenched but baseless prejudice against our proposed algorithm.

\section{On the quantum adiabatic algorithm}
A precise statement of the algorithm can be found elsewhere~\cite{kieuFull}, but in a
few words, given a Diophantine equation our aim is to obtain (by physical means or
simulations or otherwise) the (Fock) ground state of an appropriate (and bounded
from below) Hamiltonian $H_P$ carrying the information of the input Diophantine polynomial.
We will achieve that by starting with yet another easily obtained ground state of another
Hamiltonian $H_I$ and by adiabatically deforming the Hamiltonian $H_I$ in time to the one
carrying the Diophantine input above.  Essential ingredients of the algorithm are:\\
- \underline{The quantum adiabatic theorem (QAT)} (for an unbounded Hamiltonian in a
dimensionally infinite space) will ensure that, provided the adiabaticity and other
conditions are satisfied, the initial ground state will turn into the sought-after
ground state of $H_P$ with {\it high probability}.  Relevant to this point is the
paper by Tsirelson~\cite{Tsirelson} criticising our algorithm when it first came out in 2001.
This reference has not been published but somehow has been selectively cited by some as 
an evidence against our algorithm!  Those citing, intentionally or not, have either
ignored or missed out our reply to Tsirelson~\cite{kieuReply} which was posted only 3 days 
later on the same arXiv.  In that reply, we clearly pointed out that Tsirelson's
arguments were simply wrong: Had they been right, the QAT, and not just our algorithm, 
would have been mathematically wrong for all those years!\\
- \underline{No level crossing} is necessary to obtain the adiabaticity condition 
for the QAT with 
a {\em finite} rate of Hamiltonian deformation.  We have employed certain mathematical 
theorems and a gauge-like symmetry for the class of time-dependent Hamiltonians of the 
algorithm to show~\cite{kieuFull} that there is indeed no level crossing.\\
- \underline{The identification of the final ground state} is necessary because the QAT
is a nonconstructive theorem and does not tell us how the final probability of obtaining the
ground state is approached as a function of the inverse of the rate of change of the 
Hamiltonian deformation.  Such final probability certainly does not increase monotonically
but varies for different Diophantine equations in a complicated manner.  To be an
algorithm for Hilbert's tenth we need a {\em single and universal} criterion, applicable to 
any Diophantine equation, to identify the sought-after ground state.  We have 
shown~\cite{kieuFull} that it is only the final ground state that can be obtained with a 
measurement probability {\em more than 1/2} with our algorithm.  Such a probability is our
identification criterion for the ground state.  (The proof has now been
extended to an infinite number of energy levels, not just the two levels of the ground
state and the next excited state, and will appear elsewhere.)  These analytical results
have also been supported by numerical simulations~\cite{kieuOrlando}.\\
- \underline{The probabilistic nature} is inevitable for our quantum algorithm~\cite{kieuFull}, 
either in determining the identification probability through relative frequencies in physical 
measurements or in extrapolating to zero-size time step in some numerical 
simulations~\cite{kieuOrlando}.  We will come back to this probabilistic nature 
in a section below.

Further modification and extension of our work has also appeared in the 
literature~\cite{Sicard1,Sicard2}.

\section{Exploration of the infinite in a finite time?}
The recursive noncomputability of Hilbert's tenth problem lies in the fact that a systematic
substitution of positive integers (in some increasing order of magnitude, say) into a given 
Diophantine equation could only terminate if the equation has a solution; otherwise, the 
substitution would go on indefinitely without any termination point.  (For this
reason, the problem is sometimes also termed semi-computable, as we do not have a general 
method to determine when the equation has {\em no} solution.)  On the other hand, our quantum 
algorithm apparently is somehow able to ``explore the infinite (of the whole domain of positive 
integers) in a finite time!"  How can that be?  And some have even used this as an indication,
which is misleading, that there must be something wrong with our algorithm!

The logic behind all this is that Hilbert's tenth problem belongs to the class of 
{\em finitely refutable mathematical problems}~\cite{CaludeBook}.  That is, for any given
Diophantine equation it only requires a substitution up to some positive integer to 
determine whether it has any solution or not, even in the case of {\em no} solution:
if the equation has no solution within a certain {\em finite} domain of positive integers, 
it will
not have a solution anywhere else in the whole infinite domain!  The noncomputability of Hilbert's
tenth problem is precisely because we do not have any universal recursive method to determine 
this finite decisive domain for every Diophantine equation.  In contradistinction to
the recursive mathematics, quantum mechanics can give us the means to determine such 
finite decisive domain (in order to make some conclusion in the infinite domain) 
through the (energetically) ground state~\cite{kieuReply}. The finiteness 
of such a domain is encoded and reflected accordingly in the finiteness of the algorithm evolution 
time and in the finiteness of the energy and occupation number of the final ground state.  
This is how the paradoxical power of our quantum algorithm can be understood.
(In fact, the quantum algorithm provides an alternative proof for the finitely refutable 
character of Hilbert's tenth and related problems.)

\section{What about Cantor's diagonal arguments?}
Our quantum adiabatic algorithm is probabilistic in the sense that
it can produce a result with a probability, which can be made arbitrarily 
high, of being the correct result.  As such, it has a non-zero probability, 
even though it can be made arbitrarily small, {\em of being incorrect!}  
Thus, in a way, in order to compute the noncomputable or decide the undecidable 
with our algorithm, we will have to allow for the possibility of being wrong, even 
though we can reduce this chance (at a cost).
It is this probabilistic nature of the algorithm that renders it outside the
jurisdiction of Cantor's diagonal arguments~\cite{OrdKieuBJPS} employed in the 
noncomputability proof of the Turing halting problem (and hence of Hilbert's tenth 
problem)~\cite{MIT,kieuFull}.  Indeed, the discoverers of noncomputability and 
of incompleteness in Mathematics were very much aware of this power of
(probabilistic) flexibility, as reflected in their own statements~\cite{Penrose},\\
- {\bf G\"odel}: {\em `` ... it remains possible that there may exist (and even be 
empirically discovered) a theorem-proving machine which in fact is equivalent to 
mathematical intuition, but cannot be proved to be so, nor can be proved to yield 
only correct theorems of finitary number theory."}\\
- {\bf Turing}: {\em ``... if a machine is expected to be infallible, it cannot be 
intelligent. There are several theorems which say almost exactly that.  But these 
theorems say nothing about how much intelligence may be displayed if a machine makes 
no pretence at infallibility."}

Probabilistic computation may also be more powerful than Turing deterministic 
computation in yet a different way~\cite{OrdKieuCoins}, contrary to the often 
misquoted statement that the two are equivalent in terms of computability.

\section{Hypercomputation and quantum mechanics}
\subsection{A class of noncomputable linear Schr\"odinger equations}
Pour-El and Richards have shown that~\cite{PourElRichards}, surprisingly, 
the solution at a finite time of the linear wave equation in 3 dimensions
can be {\em noncomputable}, even with some computable initial condition!
This result not only shows the limitation of Turing computability even
with linear and supposedly simple differential equations like the well-known
wave equations, but also implies a hypothetical physical hypercomputation:
starting a controlled wave propagation with a precisely prepared initial 
condition and then measuring the wave configuration at a given time later
to obtain (compute) the otherwise noncomputable.
This would work provided that (i) the physical wave propagation so performed
is governed by the mathematical wave equation; and (ii) the measurement of
the wave configuration could be done with infinite precision.

Our quantum algorithm not only implies the noncomputability
of the solution of the Schr\"odinger equation with a special class of 
time-dependent Hamiltonians 
(associated with Hilbert's tenth problem) \cite{kieuPRS}; but also 
provides the procedure, physical or otherwise, 
for computing the Turing noncomputable, albeit in a probabilistic manner.  
And thanks to the flexibility of this probabilistic character, we may 
not require an infinitely precise measurement (but see below) -- but still have 
to assume that the mathematics of quantum mechanics underlies any physical 
implementation of the algorithm, unless the algorithm is simulated on Turing 
machines.

\subsection{(Probabilistic) hypercomputation and (quantum) randomness}
Beside ours, there are also other appeals to quantum mechanics as the possible
evidence and/or resource for hypercomputation~\cite{Stannett,CaludePavlov}.  
Postulated in quantum mechanics is the inherent and irreducible randomness; 
and true randomness is outside the Turing computability~\cite{Chaitin} 
and thus belongs to the domain of hypercomputation, and so does probabilistic
computation in general~\cite{OrdKieuCoins}.  Turing machines are not capable of 
generating truly random numbers, but only pseudo-random output with some finite 
and recursive algorithms.  Hypercomputation beyond the Turing computability is
thus not so mythical.  At least everyone agrees that random number generation is
a kind of hypercomputation, albeit a very special kind.

The problem is whether and how randomness can be harnessed 
for more interesting computation rather than just being random in itself.  Each series
of random numbers generated by a series of quantum measurements is different from 
each other and is not reproducible (being random by the very definition).  What 
reproducible is not the outcomes but more often is the probabilities for the outcomes.

We have pointed out how {\em some} hypercomputation could be performed with the help
of a (quantum) probability which is a noncomputable real number~\cite{OrdKieuCoins}.  
The kind of problems that can be solved by this hypercomputation 
depends on the properties of that real number.
Can we make direct use of a particular (irreproducible) series of random numbers 
generated quantum mechanically or otherwise? 
We hope to report these findings elsewhere~\cite{kieuCalude} .

\section{Physical implementation and simulations on classical computers}
Probabilistic computation may be more powerful than Turing computation
provided a suitable probability measure can be defined.  The systematic
substitution of positive integers into a Diophantine equation, for example,
does not lend itself to a definable probability measure since the cardinality 
of any subset used in the substitution is always finite while
that of all the positive integers is not.  (Recall that there is no
recursive way to determine the finite decisive subsets, even when we know 
that the problem is finitely refutable.  Were there a way, the problem 
would not have been Turing noncomputable.)

Our algorithm, on the other hand, possesses naturally defined probability measures
through the use of quantum mechanics.  In a physical implementation of the algorithm,
the probability comes from the weak law of large numbers in determining some other 
quantum probabilities through the relative frequencies obtained from 
repetitive measurements~\cite{kieuFull}.  This determination does not
require measurements of infinite precision.  However, there have been some
concerns~\cite{coeff} that infinite precision is still required in 
physically setting up the various integer parameters in the time-dependent 
quantum Hamiltonians.  While the issue deserves further investigations as surely
any systematic errors in the Hamiltonians would be fatal, we still are not
convinced that such integer parameters cannot be satisfactorily set up.  In
particular, we would like to understand the effects of statistical (as opposed to 
systematic) errors on the statistical behaviour of the spectrum of our adiabatic 
Hamiltonians.

However, physical implementation is not the only way to carry out the algorithm.
The algorithm could also be simulated on Turing machines~\cite{kieuOrlando}.  
Then how does probability come into such simulations?  It comes in under the 
necessary extrapolation of the simulation time steps to zero sizes, which is 
essentially probabilistic~\cite{kieuII}.  The probability measures here
are different from those in the physical implementation but, on the other hand, we 
do not have the above problem associated with the integer parameters in our 
Hamiltonians.

\section{Concluding remarks}
We have listed and dealt with some of the concerns and objections against our quantum
adiabatic algorithm for the Turing noncomputable tenth problem of Hilbert.
Most of these objections (none of which actually appears in print) simply root in the belief, 
and no more than a belief, 
that there  must be something wrong with the algorithm because it claims to be able
to compute the very problem that has been mathematically (and recursively)
proven to be noncomputable!  Closer inspection, however, shows that such a belief is simply 
false as all the noncomputability proof is only valid within a certain framework, outside 
of which the quantum algorithm operates and hence entails no contradiction with the 
known and proven facts.

So, what can we expect from the algorithm?  Once and if implemented, it could give an
answer, {\em with any pre-determined probability}, with respect to the existence of 
solution of any given Diophantine equation.  The probability can be raised 
arbitrarily without bounds; but the higher it is, the higher the cost of time and
resources it takes.  The other caveat is that the time it takes for a {\em successful} 
application of the algorithm is not known beforehand but only as an end product, even though 
it is always finite.  (This successful time is not the time we fix {\em a priori} for each 
run of the algorithm -- each run of the algorithm always has an end point.  We must then 
keep increasing this running time until successful.)

\section*{Acknowledgements}
I am indebted to many of my colleagues 
for the benefit I have derived through the communication and
correspondence with them.  I also wish to acknowledge in 
particular the on-going support from Peter Hannaford and Alan Head.

\end{document}